\documentclass[twocolumn,twoside,slac_two]{revtex4}

\usepackage{graphicx}
\usepackage{fancyhdr}
\usepackage{amsmath}
\def\enugam     {\ensuremath{B^+\to e^+\nu_e\gamma}\xspace }
\def\mnugam	{\ensuremath{B^+\to\mu^+\nu_{\mu}\gamma}\xspace }
\def\lnugam	{\ensuremath{B^+\to\ell^+\nu_{\ell}\gamma}\xspace }
\def\pilnu	{\ensuremath{B^+\to\pi^0\ell^+\nu_{\ell}}\xspace } 
\def\etalnu	{\ensuremath{B^+\to\eta\ell^+\nu_{\ell}}\xspace } 
\def\Xlnu	{\ensuremath{B^+\to X_u^0 \ell^+\nu_{\ell}}\xspace }  
\def\Btag	{\ensuremath{B_{\rm tag}}\xspace }  
\def\Bsig	{\ensuremath{B_{\rm sig}}\xspace } 
\def\nuMass	{\ensuremath{m_{\nu}^2}\xspace }  
\def\Npeak	{\ensuremath{N_{\ell}^{\rm peak}}\xspace }  
\def\Ncomb	{\ensuremath{N_{\ell}^{\rm comb}}\xspace }  
\def\Nbkg	{\ensuremath{N_{\ell}^{\rm bkg}}\xspace }  
\def\eff	{\ensuremath{\varepsilon_{\ell}^{\rm sig}}\xspace } 

\RequirePackage{xspace}
\usepackage{relsize}
\def\babar{\mbox{\slshape B\kern-0.1em{\smaller A}\kern-0.1em
    B\kern-0.1em{\smaller A\kern-0.2em R}}}
\def\Bp         {\ensuremath{B^+}\xspace}

\def\ellp       {\ensuremath{\ell^+}\xspace}

\def\nul        {\ensuremath{\nu_\ell}\xspace}
\def\g     {\ensuremath{\gamma}\xspace}
\def\piz   {\ensuremath{\pi^0}\xspace}
 
\def\Bbar    {\kern 0.18em\overline{\kern -0.18em B}{}\xspace}
\def\BB      {\ensuremath{B\Bbar}\xspace} 
\def\BpBm    {\ensuremath{B^+ {\kern -0.16em B^-}}\xspace}
\mathchardef\Upsilon="7107
\def\FourS {\ensuremath{\Upsilon{(4S)}}\xspace}
\def\BR         {{\ensuremath{\cal B}\xspace}}
\def\mes        {\mbox{$m_{\rm ES}$}\xspace}
\newcommand{\gev}{\ensuremath{\mathrm{\,Ge\kern -0.1em V}}\xspace}
\newcommand{\mev}{\ensuremath{\mathrm{\,Me\kern -0.1em V}}\xspace}
\newcommand{\gevc}{\ensuremath{{\mathrm{\,Ge\kern -0.1em V\!/}c}}\xspace}
\newcommand{\mevc}{\ensuremath{{\mathrm{\,Me\kern -0.1em V\!/}c}}\xspace}
\newcommand{\gevcc}{\ensuremath{{\mathrm{\,Ge\kern -0.1em V\!/}c^2}}\xspace}
\newcommand{\mevcc}{\ensuremath{{\mathrm{\,Me\kern -0.1em V\!/}c^2}}\xspace}
\def\invfb   {\ensuremath{\mbox{\,fb}^{-1}}\xspace}
\def\ps         {\ensuremath{{\rm \,ps}}\xspace}

\begin{document}

\title{Model-Independent Results for the Decay \boldmath{\lnugam} at \babar}

%

\author{D.\ M.\ Lindemann \emph{(On\ behalf\ of\ the\ \babar\ Collaboration)}}
\affiliation{Department of Physics, McGill University, Montr\'{e}al, Qu\'{e}bec, Canada H3A 2T8}

\begin{abstract}
We present a search for the radiative leptonic decays \enugam and \mnugam using data collected by the \babar\ detector at the PEP-II $B$ factory. We fully reconstruct the hadronic decay of one of the $B$ mesons in $\FourS\to\BpBm$ and then search for evidence of the signal decay within the rest of the event. This method provides clean kinematic information on the signal's missing energy and high momentum photon and lepton, and allows for a model-independent analysis of this decay. Using a data sample of 465 million $B$-meson pairs, we obtain sensitivity to branching fractions of the same order as predicted by the Standard Model.  We report a model-independent branching fraction upper limit of $\BR(\lnugam)<15.6\times10^{-6}$ ($\ell=e~ {\rm or}~\mu$) at the 90\% confidence level.
\end{abstract}

\maketitle

\thispagestyle{fancy}


\section{Introduction}
The leptonic decay \lnugam, where $\ell=e~ {\rm or}~\mu$, proceeds via an annihilation of $b$ and $u$ quarks into a virtual $W^+$ boson with the radiation of a photon.\footnote{Charge conjugate modes are included implicitly throughout this paper.}  Leptonic decays can provide clean theoretical predictions of Standard Model (SM) parameters without the QCD-based uncertainties arising from hadrons in the final state.  The purely leptonic decay $B^+\to\ell^+\nu_{\ell}$, which offers a clean prediction of the $B$-meson decay constant $f_B$, is helicity suppressed, having a branching fraction that is proportional to the square of the lepton mass.  Although the radiative mode is additionally suppressed by a factor $\alpha_{\rm em}$, the presence of the photon can remove the helicity suppression by affecting the coupling of the spin-0 $B$ meson to the spin-1 $W^{\pm}$ boson, possibly through an intermediate off-shell state \cite{ref:burdman}.  The branching fraction of \lnugam is predicted in the SM to be independent of the lepton type and of order $10^{-6}$, making it potentially accessible at current and future $B$ factories such as \babar\ and SuperB \cite{ref:superB} respectively.  The most stringent published limits are from the CLEO collaboration with $\BR(\enugam) < 2.0\times 10^{-4}$ and $\BR(\mnugam) < 5.2 \times 10^{-5}$ at the 90\% confidence level (CL) \cite{ref:cleo}.

The branching fraction of the signal decay can be written as \cite{ref:KPY}:
\begin{equation}
	\label{lnugamBF}
	\BR(B^+\!\!\to\ell^+\nu_{\ell}\gamma) = \frac{\alpha G_F^2}{288\pi^2}|V_{ub}|^2f^2_Bm_B^5\tau_B\left(\frac{Q_u}{\lambda_B}-\frac{Q_b}{m_b}\right)^2
\end{equation}
where $G_F$ is the Fermi constant, $V_{ub}$ is the Cabibbo-Kobayashi-Maskawa (CKM) matrix element describing the coupling of $b$ and $u$ quarks, $m_B$ and $\tau_B$ are the $B$-meson mass and lifetime respectively, $Q_i$ is the quark charge, and ${\lambda}_B$ is the first inverse moment of the light-cone $B$-meson wave function. This last parameter plays an important role in proving QCD factorization \cite{ref:Lunghi}.  It also enters into calculations of the ${B\to\pi}$ form factor at zero momentum transfer and the branching fractions of two-body hadronic $B$-meson decays such as ${B\to D\pi}$ and ${B\to \pi \pi}$ \cite{ref:yaouanc}, the latter being a benchmark channel for measuring the angle $\alpha$ of the CKM Unitarity Triangle.  Typically assumed to be of order $\Lambda_{QCD}$, at a few hundred MeV, ${\lambda}_B$ currently suffers from significant theoretical uncertainty \cite{ref:ball}.  A measurement of \BR(\lnugam) can provide a clean prediction of this important parameter.  In addition, since \lnugam is a possible background for $B^+\to\ell^+\nu_{\ell}$, a measurement of \BR(\lnugam) over the full photon energy spectrum is needed for an accurate measurement of \BR($B^+\to\ell^+\nu_{\ell}$)  \cite{ref:kou}, which would improve predictions of $f_B$.

\section{Analysis}
This analysis uses data taken from the \babar\ experiment \cite{ref:babar}, located at the asymmetric-energy \mbox{PEP-II} $e^+e^-$ storage rings at SLAC in California. On-peak data is produced by colliding electrons and positrons with a center-of-mass (CM) energy at the \FourS resonance of 10.58 GeV, just above the threshold for \BB production.  We use the full \babar\ dataset of $465\pm5$ million $B$-meson pairs, corresponding to an integrated luminosity of 423~\invfb.  Charged-particle momenta are measured using a tracking system comprised of a silicon vertex detector and drift chamber contained within the uniform magnetic field of a 1.5T superconducting solenoid.  Charged-particle identification is based on the energy loss in the tracking system and the Cherenkov angle in a ring-imaging Cherenkov detector.  Photon energies and electron identification are provided by a CsI(Tl) scintillating electromagnetic calorimeter (EMC).  Finally, muons are distinguished from hadrons using instrumentation embedded within the steel magnetic-flux return.  Monte Carlo (MC) simulations are used to model the detector response, determine the signal efficiency, and study the background.

\subsection{Hadronic Reconstruction}
We present the first search for \lnugam that uses an exclusive method, by fully reconstructing one $B$ meson (\Btag) via hadronic decay modes.  Unlike previous inclusive searches of \lnugam, which identified the signal lepton and photon candidates before the recoiling $B$ meson, we reconstruct the recoil $B$ meson first before searching for the signal decay. Although this technique results in a low signal efficiency (0.3\% for signal modes), and thus more statistically-limited results, it compensates by providing a highly pure sample of $B$ mesons with comparatively little non-\BB (continuum) background.  Our reliance on poorly-modeled continuum background is thus reduced, which enables a search for \lnugam that can avoid kinematic constraints in the signal selection.  Therefore, we present the first model-independent search of \BR(\lnugam) over the full kinematic range.  In addition, by reconstructing the \Btag using only detectable hadronic decay modes, the missing four-vector of the otherwise undetectable signal neutrino is fully determined.  

The event selection begins by reconstructing a $D^{(*)}$ meson using various hadronic decay modes, listed in Ref.~\cite{ref:mine}.  Additional charged tracks and neutral EMC clusters are then combined with the $D^{(*)}$ seed to form the decay ${B\to D^{(*)}X_{\rm had}}$, where $X_{\rm had}$ is a combination of neutral and charged pions and/or kaons.  $X_{\rm had}$ is chosen such that $\lvert\Delta E\rvert \equiv \lvert E_{\Btag}-\frac{\sqrt{s}}{2}\rvert<0.12$, where $\sqrt{s}$ is the total energy of the $e^+e^-$ system and $E_{\Btag}$ is the \Btag candidate energy, both in the CM frame.  If there exists more than one \Btag candidate, we choose the one with the $D^{(*)}$ and $X_{\rm had}$ mode that provides the highest purity of well-reconstructed candidates.  Finally, we require that the \Btag candidate is charged.

Since the beam energy at PEP-II is accurately known within a few MeV, we define the mass of the \Btag as $\mes=\sqrt{\frac{s}{4}-{\vec p}_{\Btag}^{~2}}$, where ${\vec p}_{\Btag}$ is the \Btag three-momentum in the CM frame.  Since the \Btag candidate should have a mass consistent with the nominal $B$-meson mass, we require $5.27<\mes<5.29\gevcc$. The \mes sideband region, defined as ${5.20<\mes<5.26}\gevcc$, is populated by combinatoric background, events coming from a $\FourS\to\BB$ or continuum decay in which the \Btag is incorrectly reconstructed.  To improve the MC estimation of the reconstruction efficiency, we normalize the MC that peaks within the \mes signal region to the data that peaks within this same region, after applying the \Btag selection criteria.  In addition, we reduce our reliance on the MC estimate of combinatoric events by estimating the non-peaking background directly from the data within the \mes sideband.

When a \FourS resonance decays to a \BB pair, the two particles are almost at rest, with a momenta of about 350\mevc in the CM frame, since their masses are about half the \FourS mass.  Therefore, they tend to decay with an isotropically symmetric topology in the CM frame.  On the other hand, lighter $q\overline q$ and $\tau^+\tau^-$ pairs, which are also produced in the $e^+e^-$ interactions, have a higher momentum in the CM frame.  Their decays tend to have a more jet-like shape with a strongly-preferred direction characterizing the event, preferentially at small angles in relation to the beam axis.  Therefore, to achieve a higher purity of $B$ mesons, we use a multivariate selector of five event-shape variables to separate \BB from continuum events.  We require:
\begin{equation}
{\cal L}_{B}\equiv \frac{\prod_i {\cal P}_{B}(x_i)}{\prod_i {\cal P}_{B}(x_i)+\prod_i {\cal P}_{q}(x_i)}>30\% ,
\end{equation}
where ${\cal P}_{B}(x_i)$ and ${\cal P}_{q}(x_i)$ are probability density functions determined from MC that describe \BB and continuum events, respectively, for the five event-shape variables $x_i$.  These variables, as explained in Ref.~\cite{ref:mine}, describe the sphericity of the event, the magnitude and directions of the thrust axes, and the direction of the \Btag momentum with respect to the beam axis.  This requirement also suppresses the poorly-modeled continuum backgrounds which can contribute to discrepancies between the MC and data. 

\subsection{Signal Selection}
After the \Btag reconstruction, we assign all the remaining charged tracks and neutral EMC clusters, to the signal $B$ meson (\Bsig), as well as any missing momentum (${\vec p}_{\rm miss}$) within the event.  We require that \Bsig has exactly one track, with a charge opposite that of the \Btag charge. In addition, this signal track must satisfy particle identification criteria of either an electron or muon and fail that of a kaon.

Because high energy electrons can emit bremsstrahlung photons, we search for signal-side clusters that have proximity to the EMC deposit of an electron candidate, both in $\theta$, the angle from the beam axis, and in $\phi$, the azimuthal angle around the beam axis.  Because the detector has a solenoid magnet through which all the particles travel, charged particles are bent in $\phi$ according to their charge, and therefore ${\Delta\phi}$ is multiplied by the lepton candidate charge.  Any cluster with a momentum vector that is separated from that of the electron-identified signal track by $|\Delta\theta|<3^{\circ}$ and ${-3^{\circ}<\Delta\phi<13^{\circ}}$ is identified as a bremsstrahlung photon.  The energy of this cluster is then used to correct the four-vector of the signal track ($p_{\ell}$) and is removed from the list of clusters.  Next, we identify the signal photon candidate, whose energy spectrum is expected to peak around 1~\gev.  We do this by searching through the remaining signal-side clusters for the one with the highest energy in the CM frame.

Although \lnugam should only have one signal-side cluster if correctly reconstructed, extra clusters within an event can be due to fragments from particle showers, low-energy clusters from the \Btag, and/or noise in the detector such as beam-related photons.  Because of the presence of extra clusters, we apply the very loose requirement that there are no more than 11 additional low energy clusters which are not otherwise used in the \Btag or signal reconstruction.  The total energy of these clusters is required to be less than 800\mev, where we only include in this sum clusters which have lab-frame energy greater than $50\mev$.

Missing energy can be due to either an undetectable particle such as a neutrino or a detectable particle that travels outside of the fiducial acceptance of the detector.  To suppress events of the latter category, we require that ${\vec p}_{\rm miss}$ points within the fiducial acceptance of the detector.

The lepton and neutrino are produced back-to-back by the virtual $W^{\pm}$ boson.  However, in the \Bsig rest frame, the angle $\theta_{\ell\nu}$ between the signal track momentum and ${\vec p}_{\rm miss}$ is affected by the release of the photon. Therefore, we require $\cos\theta_{\ell\nu}< -0.93$ in the frame recoiling from the photon.  This frame is defined as the difference between the \Bsig four-vector ($p_B$) and the signal photon candidate four-vector ($p_\g$).  Including this requirement assures that the decay is consistent with being a three-body decay, independent of when the photon was released.

\begin{figure}[h]
\includegraphics[width=85mm]{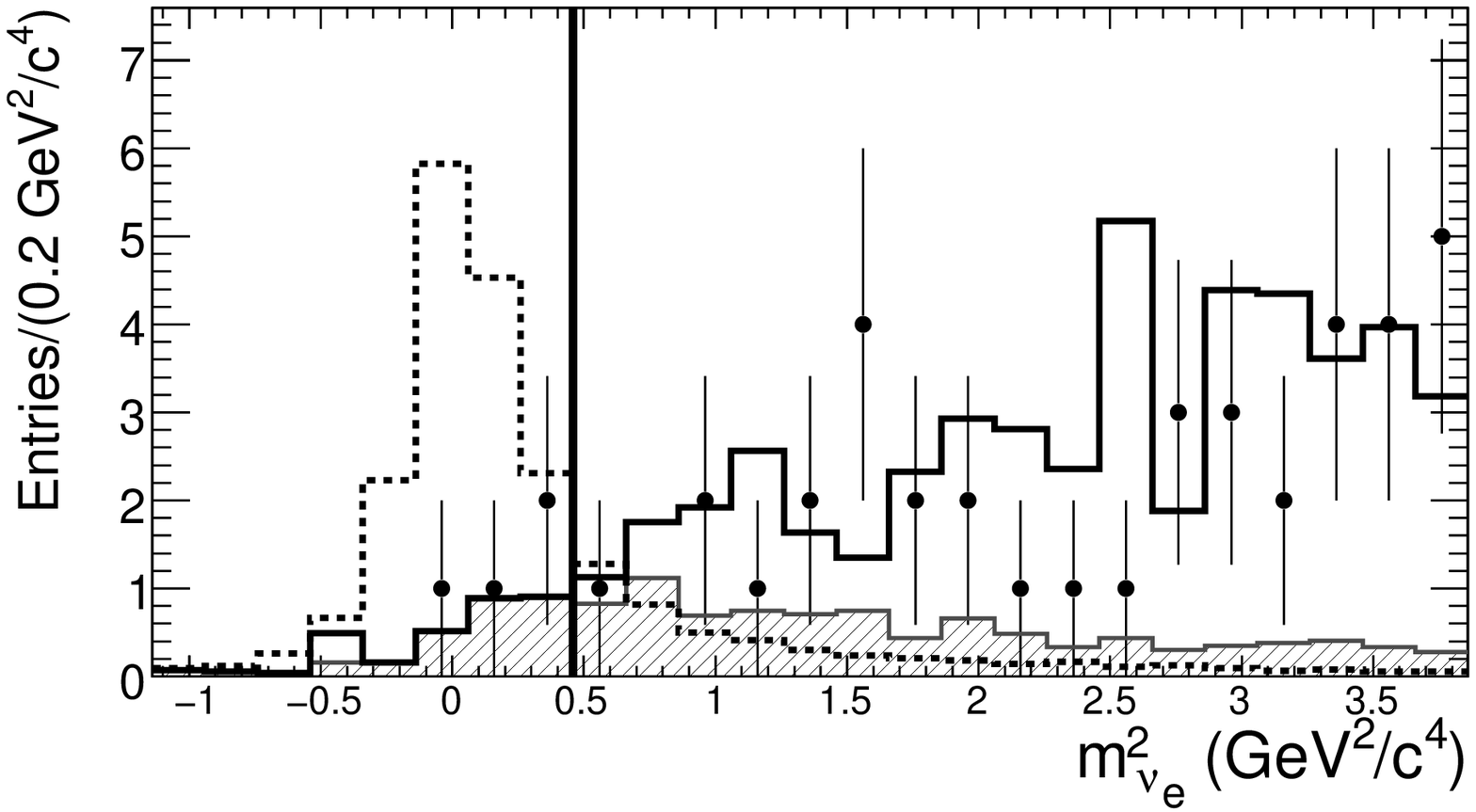}
\hfill
\includegraphics[width=85mm]{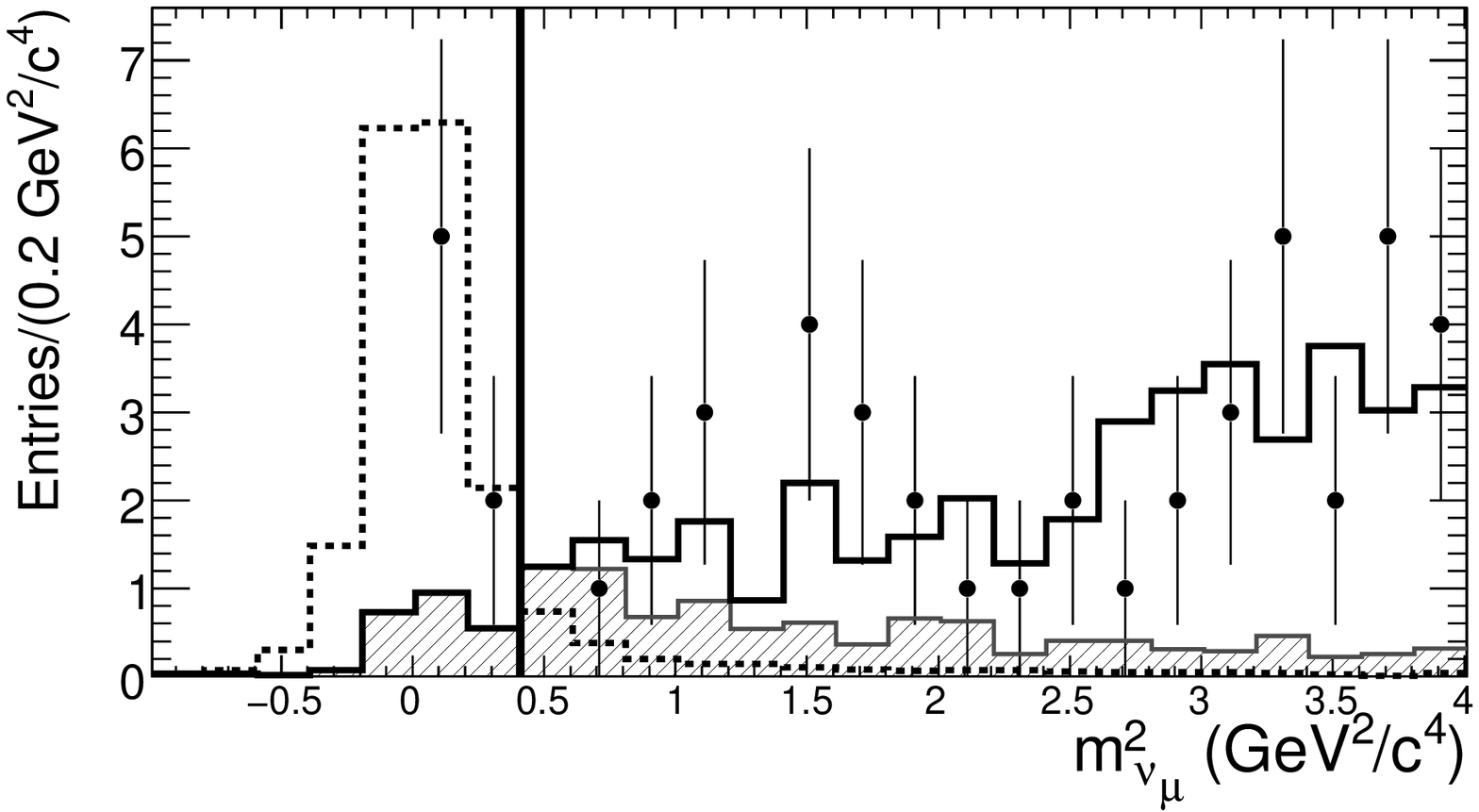}
\caption{\nuMass distribution after all selection criteria are applied, in electron (top) and muon (bottom) modes for the \mes-peaking (shaded) plus non-peaking (solid) contributions in the full background MC sample, signal MC normalized to $\BR=40\times10^{-6}$ (dashed), and data (points).  Events to the left of the vertical lines are selected.  \label{fig:nuMass}}
\end{figure}

The most discriminating variable in this analysis is a requirement on the neutrino candidate's invariant mass \nuMass, given by:
\begin{equation}
	\label{nuMass2}
	\nuMass=(p_{B}-p_{\gamma}-p_{\ell})^2 .
\end{equation}  This requirement isolates events in which the photon and lepton candidates are in kinematic agreement with a massless third daughter in the event.  Fig.~\ref{fig:nuMass} shows that the signal peaks at zero, while the the background rises with \nuMass.  The tail on the signal distribution is larger for the electron mode than the muon mode due to unrecovered bremsstrahlung photons.  The signal region for this variable is defined as ${-1<\nuMass<0.46~(0.46)\gev^2/c^4}$ for the electron (muon) mode.  These values, as with all other cut values in this analysis, were optimized using the figure of merit $\eff/(\tfrac{1}{2}n_{\sigma}+\sqrt{\Nbkg})$ \cite{ref:punzi}, where $n_{\sigma} = 1.3$, \eff is the total signal efficiency, and $\Nbkg$ is the number of expected background events.

The \nuMass and $\cos\theta_{\ell\nu}$ requirements kinematically restrict the types of events that pass.  Since light ${X_u}$ mesons often decay to a pair of photons, where $X_u$ is a neutral meson containing a $u$-quark, the topology and kinematics of a \Xlnu event can often mimic that of the signal decay.  This is particularly true if one photon is ``missing" such as through misreconstruction into the \Btag or due to a lab-frame energy that is too low to be detectable by the EMC.  Thus, the primary background that passes the signal selection are \Xlnu events with a high energy photon.  To suppress \Xlnu events in which both photon daughters are present in the signal-side clusters, we reject events in which the signal photon candidate can be combined with another cluster to form an invariant mass consistent with the \piz or $\eta$ mass.  To improve the purity of the reconstructed \piz mesons, we only use clusters that are above a given energy $E_{\gamma 2}$ in the \Bsig rest frame.  To suppress \pilnu events, we reject events in which the invariant mass is between 120 and 145\mev with $E_{\gamma 2}>30\mev$ or between 100 and 160\mev with $E_{\gamma 2}>80\mev$.  Since $\eta$ particles are much more massive and thus tend to decay into higher energy photons, we suppress \etalnu events by rejecting events with $E_{\gamma 2} > 100 MeV$ and an invariant mass between 515 and 570\mev.  Finally, we suppress $\Bp\to\omega\ellp\nul\to[\piz\g]\ellp\nul$ events by rejecting events in which the signal photon candidate combined with a \piz candidate forms an invariant mass between 115 and 145\mev.  This \piz candidate must have $E_{\gamma 2}>70\mev$ and an invariant mass between 115 and 145\mev.

Events in which the two photons from a \pilnu decay are merged into a single EMC cluster can mimic the signal kinematics exactly, since the cluster chosen as the signal photon candidate contains the full energy of the \piz.  We suppress this background using a requirement on the cluster width of the signal photon candidate.  A cluster that is the result of two merged photons from a \piz decay tends to be wider than a cluster from a single photon.  We require the lateral moment \cite{ref:latMom} to be less than $55\%$.

\subsection{Backgrounds and Uncertainties}
We define the signal branching fraction for each lepton mode $\ell$ as:
\begin{equation}
	\label{BFeq}
	\BR_{\ell}=\frac{N^{\rm obs}_{\ell} - \Nbkg}{\eff N_{B^{\pm}}}
\end{equation}
where $N^{\rm obs}_{\ell}$ is the number of observed data events within the signal region and $N_{B^{\pm}}=465\times10^6$ is the number of $B^{\pm}$ mesons in the data sample.  The branching fractions are computed using the frequentist formalism of Feldman and Cousins \cite{ref:FC}, with uncertainties on \Nbkg and \eff modeled using Gaussian distributions.  Since \BR(\lnugam) is expected to be independent of the lepton type, we also combine the branching fractions of the two modes.

The MC indicates that the selection criteria, especially the one track requirement and the \nuMass and $\cos\theta_{\ell\nu}$ restrictions on the event kinematics, removes all decay types except \Xlnu events and the occasional combinatoric event from a misreconstructed \Btag.  Thus, we split \Nbkg into two categories: peaking events (\Npeak) and combinatoric events (\Ncomb).  An event is considered peaking if it comes from a $\FourS\to\BB$ decay in which the \Btag is correctly reconstructed and hence peaks within the \mes signal region.  Since only \Xlnu events peak in this region, we improve the statistics of \Npeak by using exclusive \Xlnu MC samples.  For \Ncomb, we extrapolate the number of misreconstructed \Btag events in the \mes signal region directly from the number of data events in the \mes sideband.

The uncertainty on \Ncomb is dominated by the sideband data statistics, but it also includes a 14.6\% systematic uncertainty on the \mes combinatoric background shape.  On the other hand, since \eff and \Npeak rely on the accuracy of the MC simulations, their uncertainties arise not only from MC statistics, but also include systematic uncertainties due to how well the MC agrees with the data.  The contributions to the systematic uncertainties are listed in Table \ref{tab:sysSummary}.  The systematic uncertainty of \Npeak is dominated by a 13.6\% uncertainty in the branching fractions and form factors of various exclusive \Xlnu decays \cite{ref:mine}.  The uncertainty due to the \Btag reconstruction, which also accounts for the uncertainty in $N^{B^{\pm}}$, is determined by varying the shape and scaling of the \mes combinatoric distribution.
\begin{table}[h]
\begin{center}
	\caption{Contributions (in percent) to the systematic uncertainty on the branching fraction due to the signal efficiency \eff and the peaking-background estimate \Npeak.  The \BR(\Xlnu) source refers to branching fraction and form factor uncertainties in \Xlnu decays. \label{tab:sysSummary}} 
 	\begin{tabular}{|l|c|c|c|c|} \hline
Source of  & \multicolumn{2}{c|}{\eff} & \multicolumn{2}{c|}{\Npeak}\\ 
systematics & $e$ mode & $\mu$ mode & $e$ mode & $\mu$ mode \\ \hline \hline
\BR(\Xlnu) & -- & -- & 13.6 & 13.6  \\ \hline
\Btag reconstruction & 3.1 & 3.1 & 3.1 & 3.1 \\ \hline
Particle identification & 0.9 &1.3 & 0.9 & 1.3 \\ \hline
Track reconstruction & 0.4 & 0.4  & 0.4  & 0.4  \\ \hline
Photon reconstruction & 1.8 & 1.8 & 1.8 & 1.8 \\ \hline
${\cal L}_{B}$  & 1.4 & 1.4& 1.4& 1.4 \\ \hline
\nuMass &  0.5 &0.5 & 1.4 & 1.4  \\ \hline
 	\end{tabular}
	\end{center}
\end{table}

\section{Results}
To avoid experimenter bias, we blinded the data in the region $\nuMass<1\gev^2/c^4$ and $\mes>5.26\gevcc$ until all selection criteria, background estimates, and systematic uncertainties were finalized.  The final signal efficiencies correspond to approximately one signal event per mode, assuming a branching fraction within the SM predictions.  After unblinding, we observe 4 (7) data events within the signal region for the electron (muon) mode, compared to an expected background of 2.7 $\pm$ 0.6 (3.4 $\pm$ 0.9) events.  The branching fraction results are given in Table~\ref{tab:finalNums}.
\begin{table}[h]
	 \caption{Expected background yields $\Nbkg\!=\Ncomb+\Npeak$, signal efficiencies \eff, number of observed data events $N_{\ell}^{\rm obs}$, resulting branching fraction limits at 90\% CL, and the combined central value $\BR_{combined}$.  Uncertainties are given as statistical $\pm$ systematic. \label{tab:finalNums}}
	\begin{center}
 	\begin{tabular}{|l|c|c|} \hline
 & \enugam & \mnugam \\ \hline \hline
\raisebox{-1pt}{\Ncomb} & 0.3 $\pm$ 0.3 $\pm$ 0.1 & 1.2 $\pm$  0.6 $\pm$  0.6\\
\raisebox{-1pt}{\Npeak} & 2.4 $\pm$  0.3 $\pm$  0.4 & 2.1 $\pm$  0.3 $\pm$  0.3\\ \hline
\raisebox{-1pt}{\Nbkg} & 2.7 $\pm$  0.4 $\pm$  0.4 & 3.4 $\pm$  0.7 $\pm$  0.7\\
\raisebox{-1pt}{\eff} & (7.8 $\pm$  0.1 $\pm$  0.3)$\times10^{-4}$
			     & (8.1 $\pm$  0.1 $\pm$  0.3)$\times10^{-4}$ \\
\raisebox{-1pt}{$N_{\ell}^{\rm obs}$} & 4 & 7  \\ \hline
\raisebox{-1.5pt}{$\BR_{combined}$} 
	& \multicolumn{2}{c|}{\raisebox{-1.5pt}{$\bigl(6.5 _{-4.7}^{+7.6}$$_{-0.8}^{+2.8}\bigr)\times10^{-6}$}} \\ \hline
\raisebox{-1.5pt}{\footnotesize{Model-ind.}} & \raisebox{-1.5pt}{$< 17$$\times10^{-6}$}
           & \raisebox{-1.5pt}{$< 26$$\times10^{-6}$} \\ 
\raisebox{-1.5pt}{Limits}  & \multicolumn{2}{c|}{\raisebox{-1.5pt}{$<15.6\times10^{-6}$}} \\ \hline
 	\end{tabular}
	\end{center}
 \end{table}

\begin{figure*}[t]
\centering
\includegraphics[width=55mm]{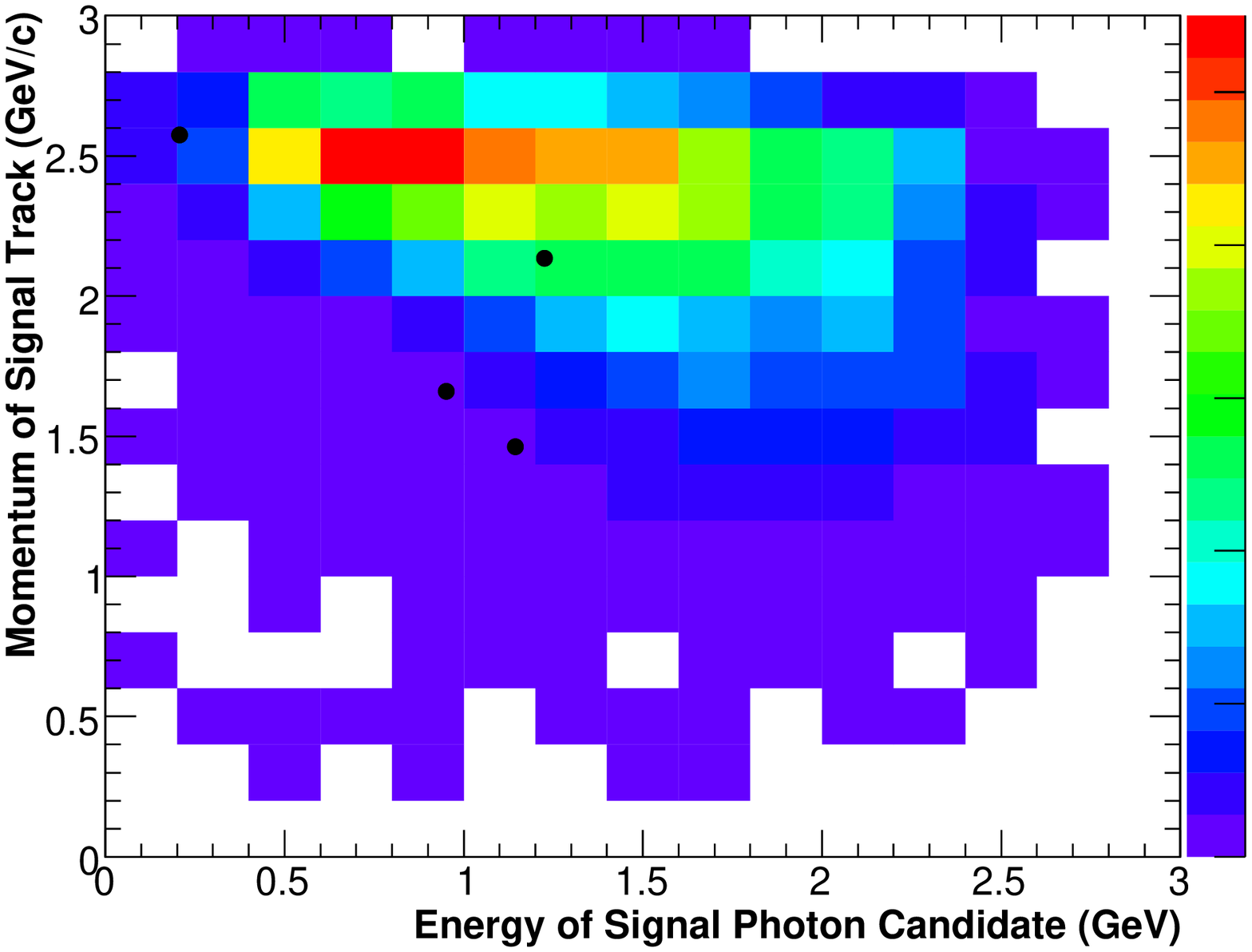}
\hfill
\includegraphics[width=55mm]{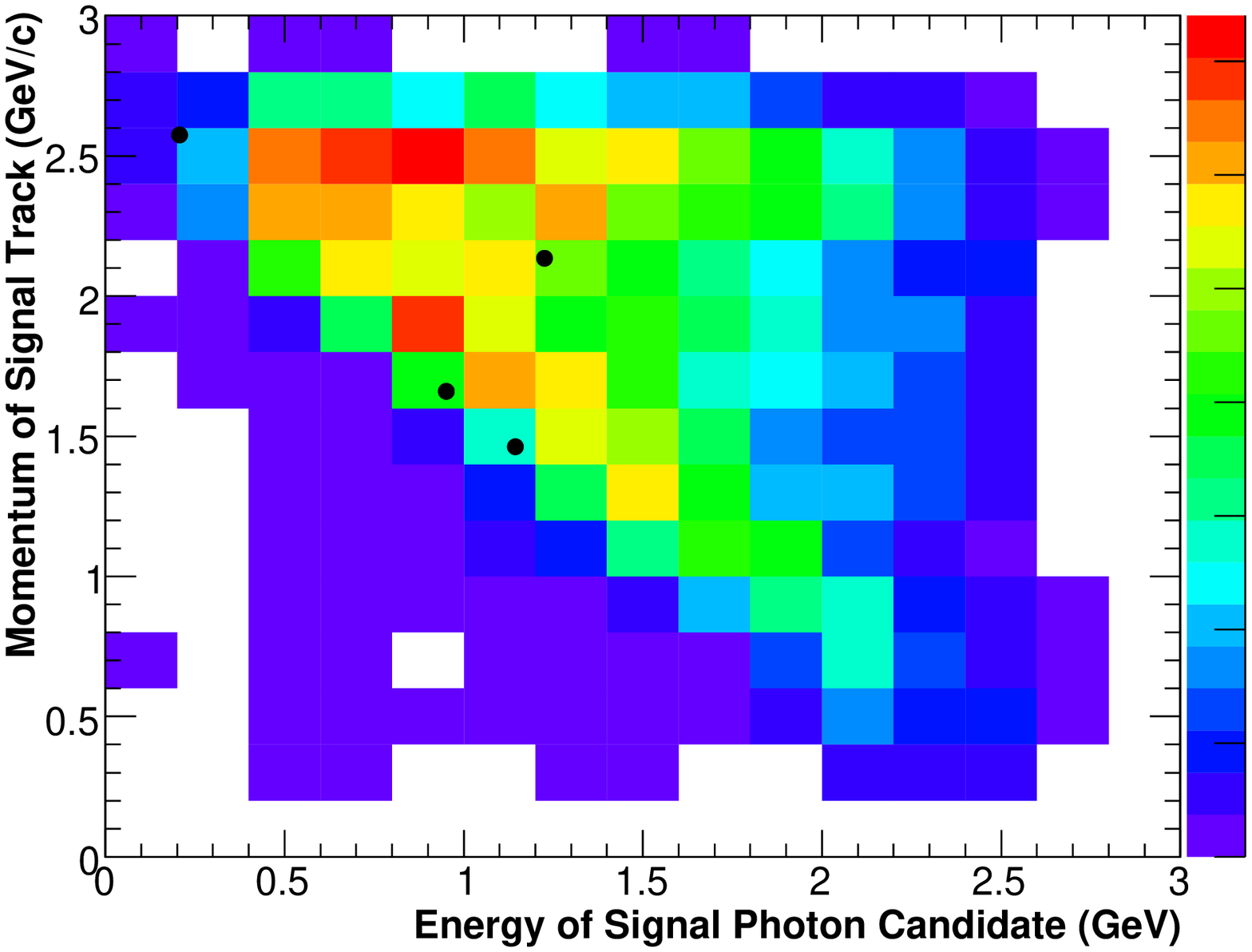}
\hfill
\includegraphics[width=55mm]{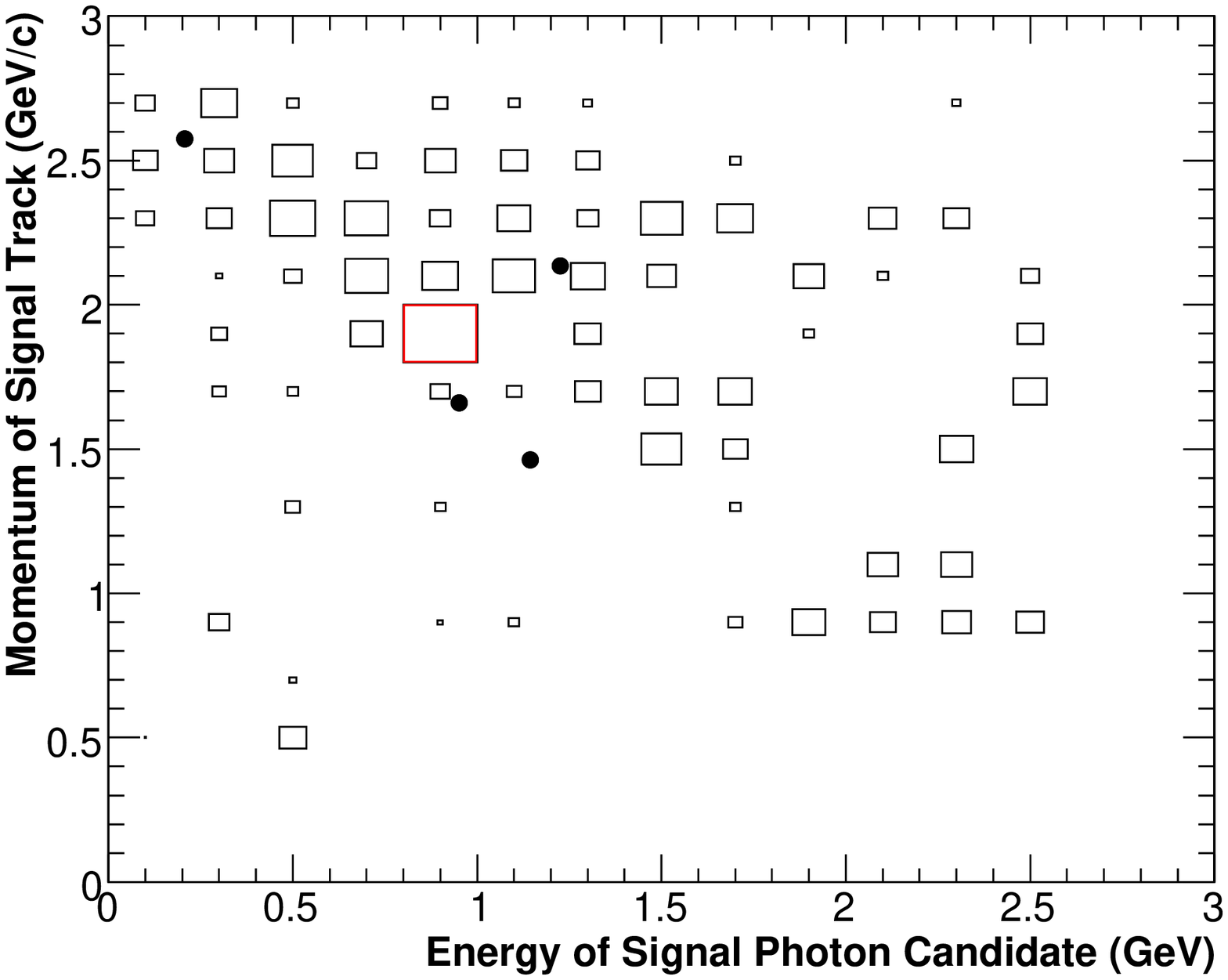}
\hfill
\includegraphics[width=55mm]{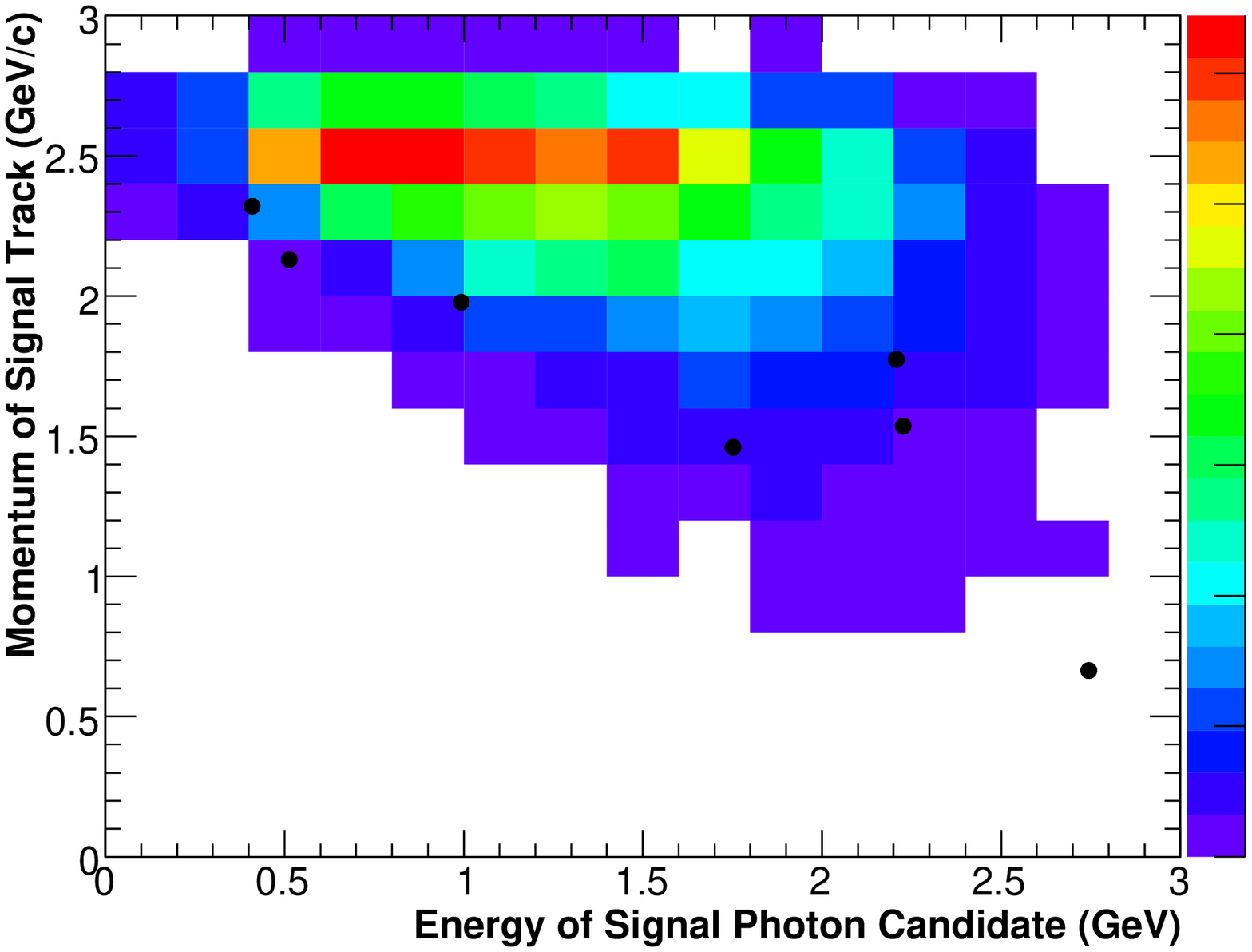}
\hfill
\includegraphics[width=55mm]{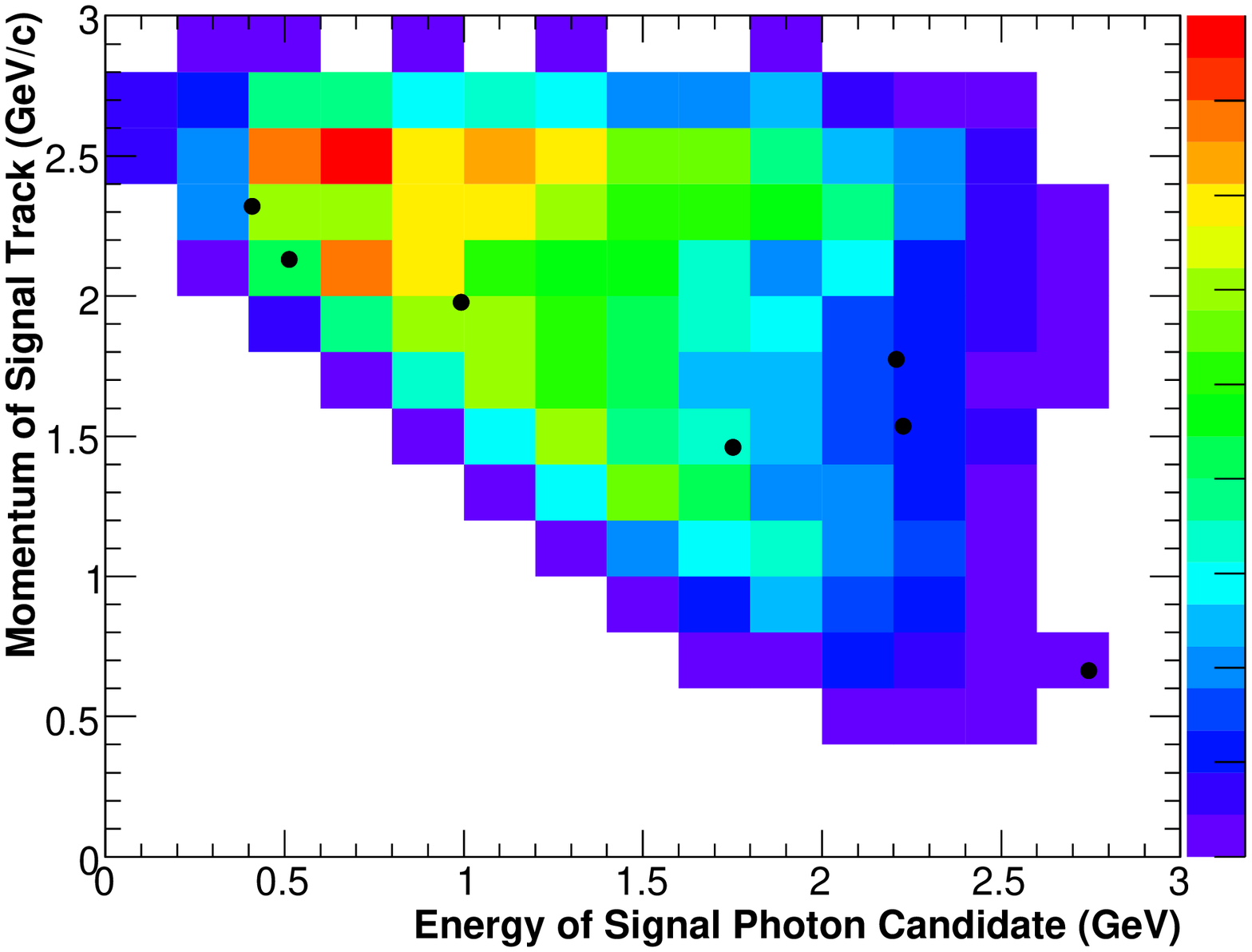}
\hfill
\includegraphics[width=55mm]{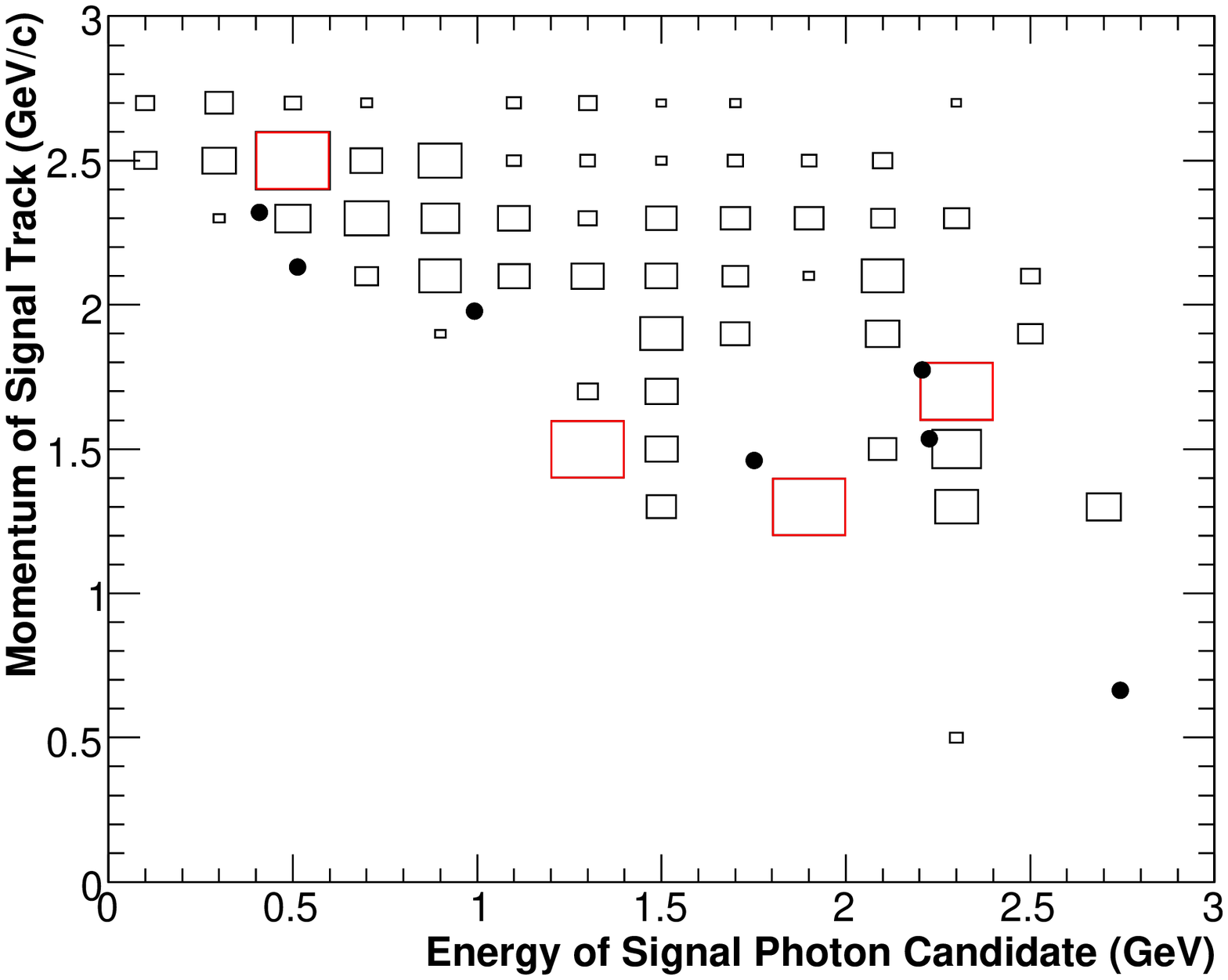}
\caption{The distribution of the signal track momentum versus the signal photon candidate energy, in the \Bsig rest frame, after all signal selection criteria is applied, for data (points), $f_A=f_V$ signal model (left), $f_A=0$ signal model (middle), and \Nbkg (right).  The electron modes are shown on the top and the muon modes are on the bottom.  The size of the box in the \Nbkg plots is proportional to the number of background events within the histogram bin.  The contribution of \Ncomb, which is determined using data events in the \mes sideband, is shown as red boxes.  \label{fig:phasespace}}
\end{figure*}

Although the effective detector and particle identification thresholds are about 20\mev for photon energy and 400 (800)\mev for electron (muon) momentum, and we apply no minimum energy requirements.  Thus, this analysis is essentially valid over the full kinematic range, as shown in Fig.~\ref{fig:phasespace}.  However, there exists theoretical uncertainty in the photon energy spectrum below $\Lambda_{\rm QCD}$ for the \lnugam decay.  Therefore, using certain theoretical techniques, the extraction of $\lambda_B$ can be improved by including a minimum energy requirement on the signal photon \cite{ref:ball}.  When the signal photon candidate energy is required to be greater than 1\gev, we observe 2 (4) data events with $\Nbkg=1.4 \pm 0.3~(2.5 \pm 1.0)$ in the electron (muon) mode.  The \eff, which is mostly uncorrelated with the photon energy spectrum, is reduced by 30\%, resulting in a partial branching fraction of $\Delta\BR(\lnugam)<14\times10^{-6}$ at 90\% CL.

The differential branching fraction versus photon energy $E_{\g}$ of \lnugam is given by:
\begin{equation}
	\label{lnugamDiffBF}
	\frac{d\Gamma}{dE_{\g}} = \frac{\alpha G_F^2}{48\pi^2}|V_{ub}|^2m_B^4\left(f_A^2(E_{\g})+f_V^2(E_{\g})\right)x(1-x)^3
\end{equation}
where ${y\equiv2E_{\g}/m_B}$.  The two form factors, $f_V$ and $f_A$, describe the vector and axial-vector contributions, respectively, to the $B\to\g$ transition.  Although \mbox{$f_A=0$} in some models \cite{ref:burdman} and was assumed in the CLEO measurement of \BR(\lnugam), most models assert $f_A=f_V$ \cite{ref:Lunghi}.  In this analysis, we use signal MC that is generated based on the tree-level hadronic matrix element for \lnugam as described by Ref.~\cite{ref:KPY}, using a minimum photon energy of 350 MeV.  We use both form-factor models for our signal MC to evaluate the impact of the decay model on the signal selection efficiency and to ensure model-independency.  We determine \eff using the $f_A=f_V$ signal model, but because our analysis is independent of the decay kinematics, the $f_A=0$ model yields consistent \eff values.

\begin{figure*}[t]
\centering
\includegraphics[width=55mm]{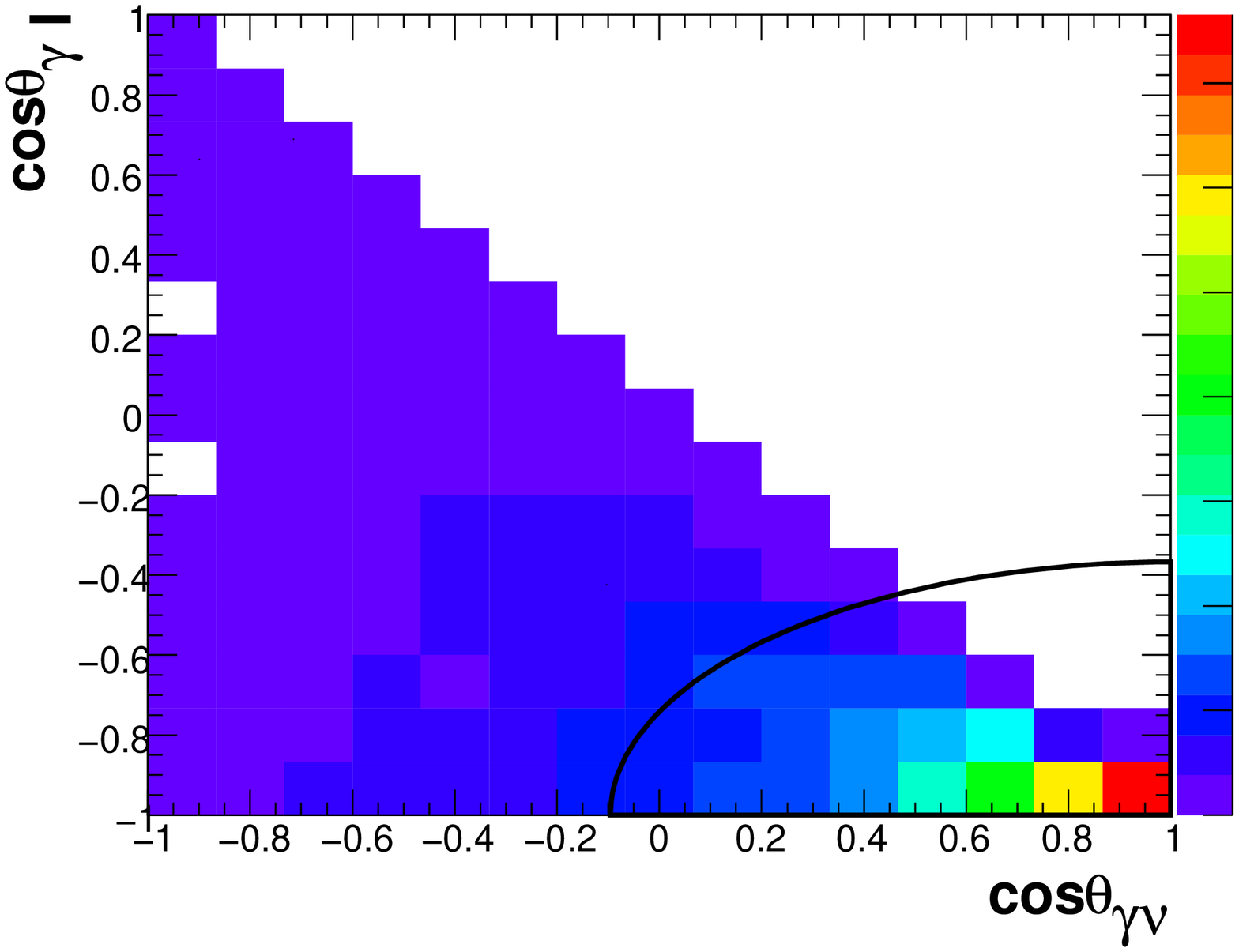}
\hfill
\includegraphics[width=55mm]{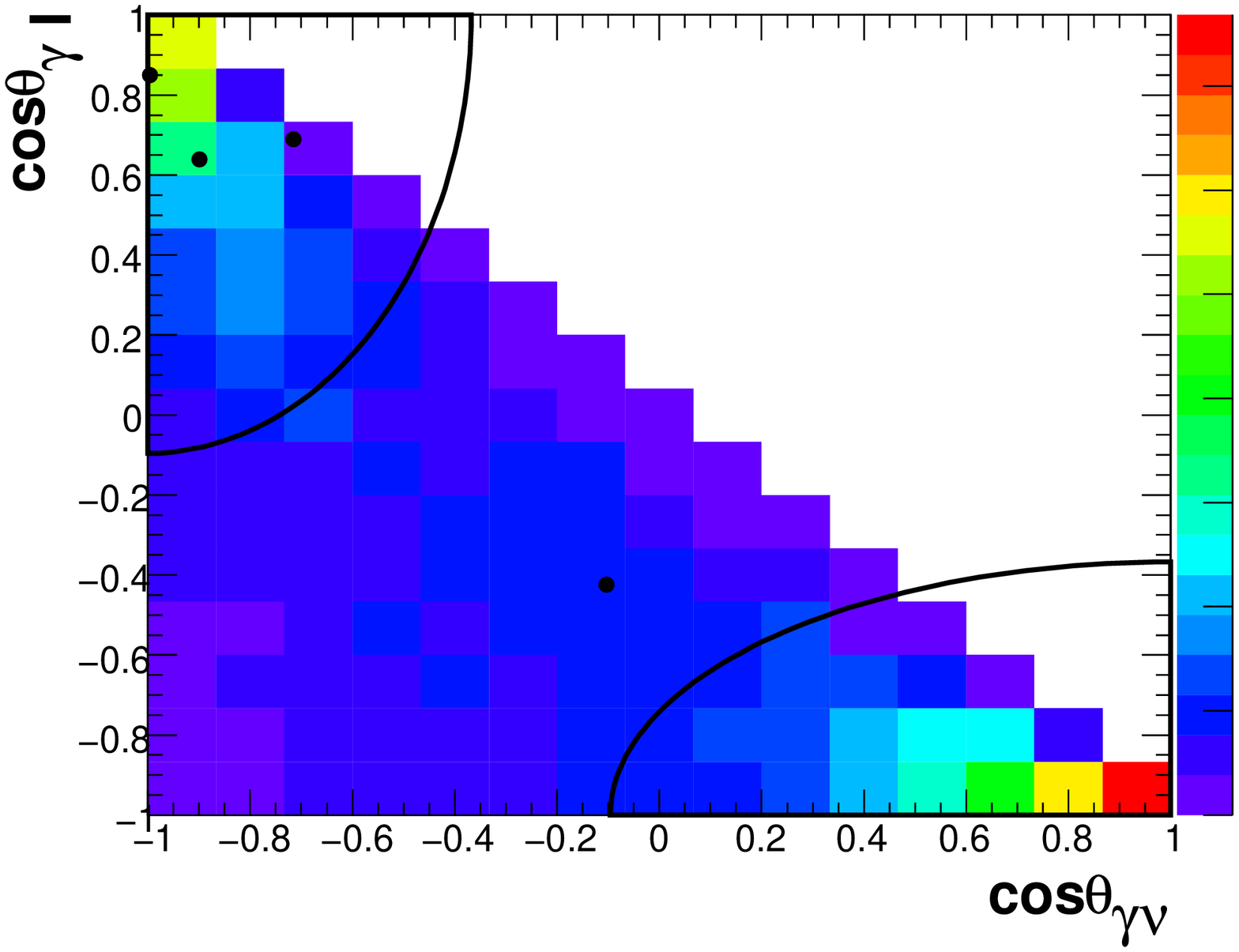}
\hfill
\includegraphics[width=55mm]{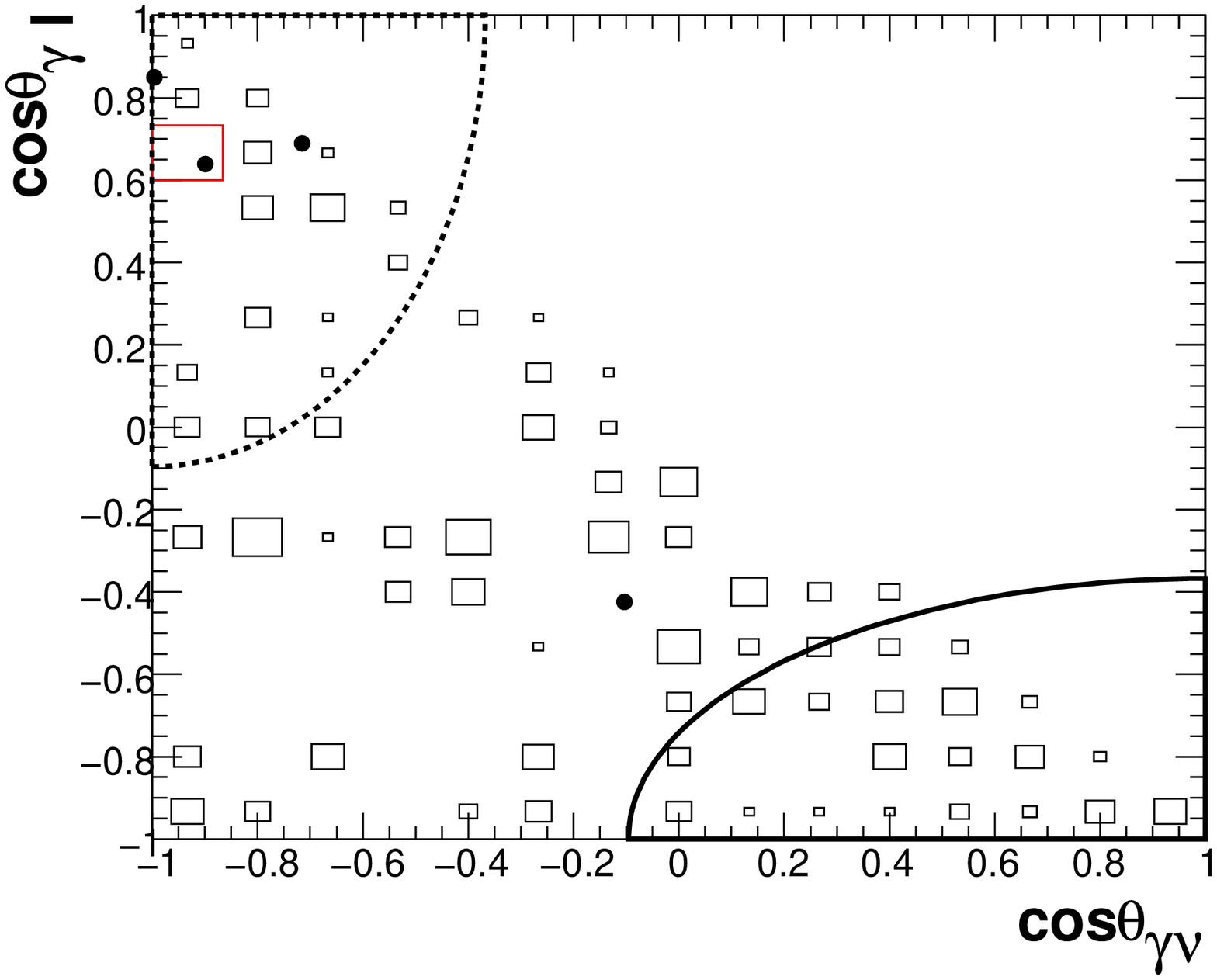}
\hfill
\includegraphics[width=55mm]{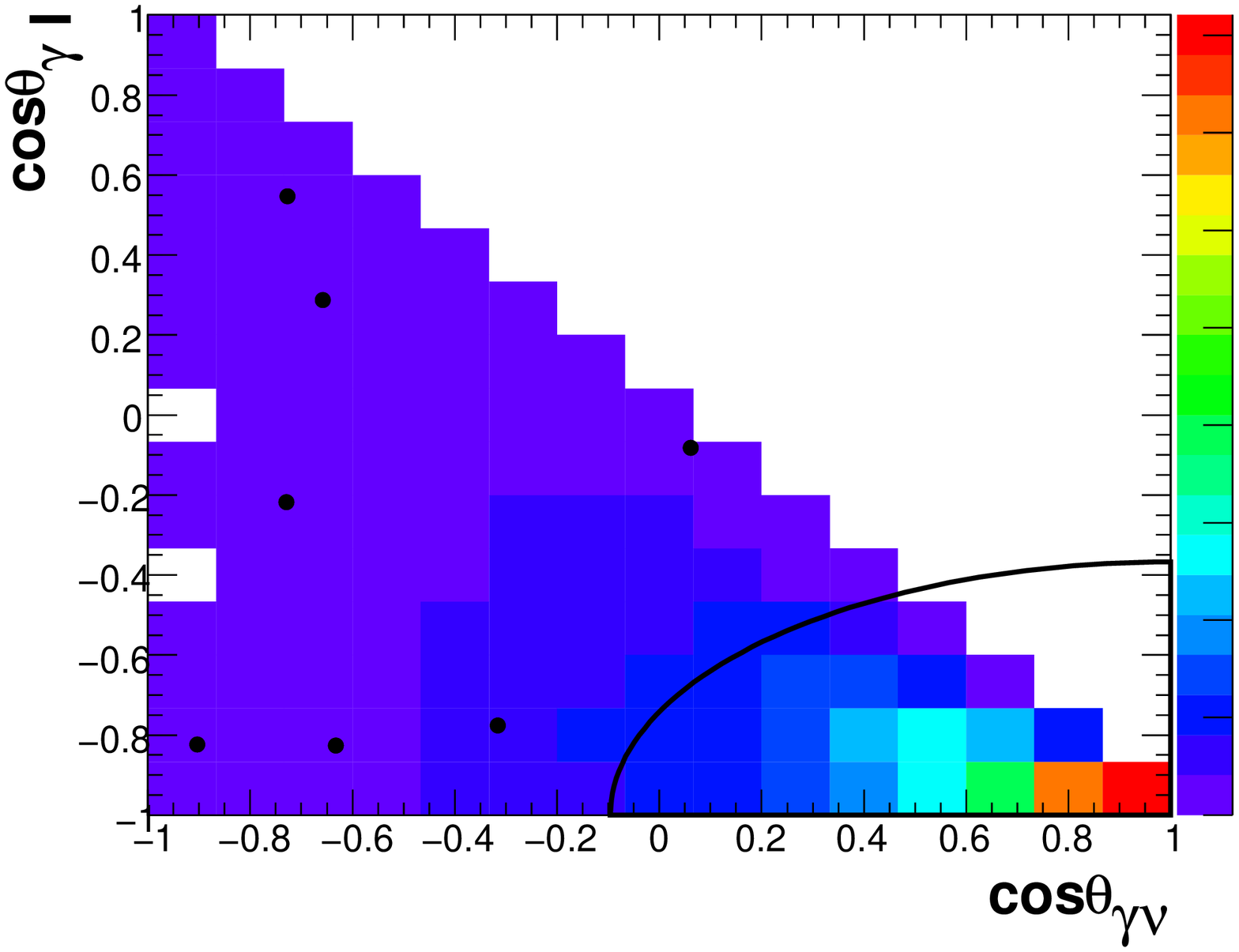}
\hfill
\includegraphics[width=55mm]{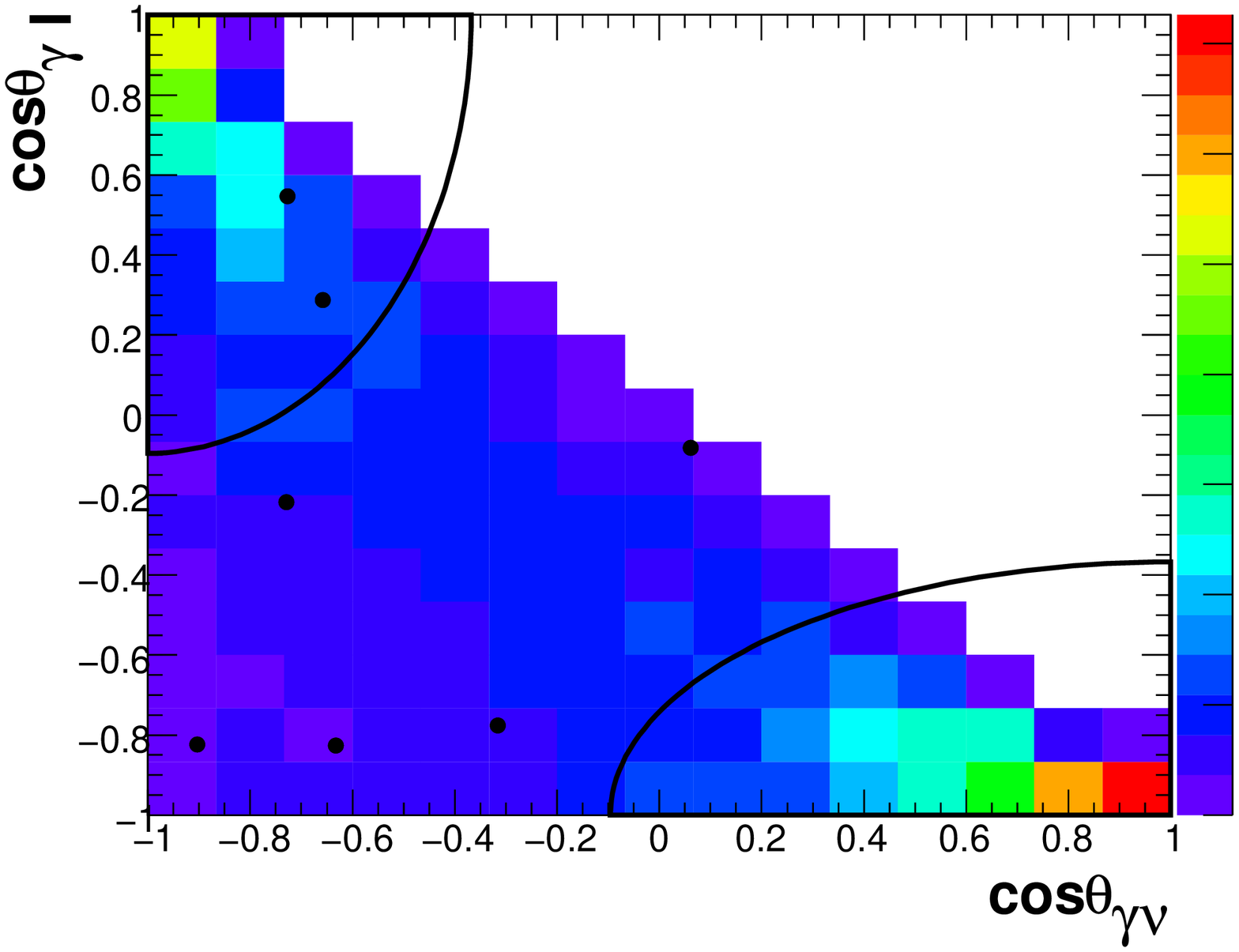}
\hfill
\includegraphics[width=55mm]{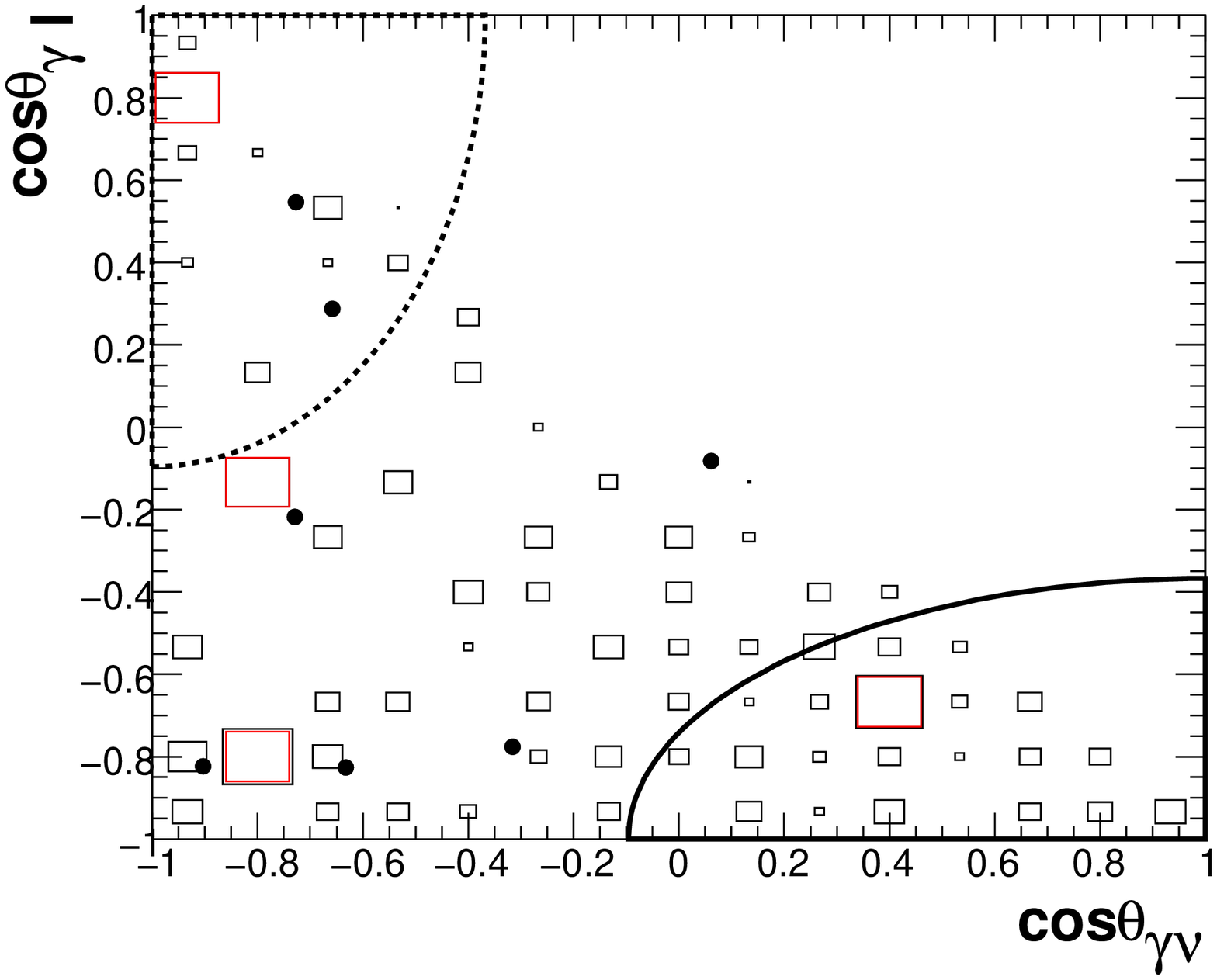}
\caption{The correlation between $\cos\theta_{\g\ell}$ and $\cos\theta_{\g\nu}$, in the \Bsig rest frame, after all signal selection criteria is applied, for data (points), $f_A=f_V$ signal model (left), $f_A=0$ signal model (middle), and \Nbkg (right).  The electron modes are shown on the top and the muon modes are on the bottom.  The size of the box in the \Nbkg plots is proportional to the number of background events within the histogram bin.  The contribution of \Ncomb, which is determined using data events in the \mes sideband, is shown as red boxes.  The black arcs indicate the cut-off values for the model-specific requirements. \label{fig:angles}}
\end{figure*}

We also determine branching fraction limits that are dependent on the signal model by introducing a kinematic requirement on the angles between the three daughter particles of the signal decay.  We use the quantities $\cos\theta_{\g\ell}$ and $\cos\theta_{\g\nu}$, where $\theta_{\g\ell}$ is the angle between the photon candidate and signal track momenta, and $\theta_{\g\nu}$ is the angle between the photon candidate momentum and ${\vec p}_{\rm miss}$, both in the \Bsig rest frame.  As seen in Fig. \ref{fig:angles}, the photon is emitted preferentially back-to-back with the lepton in the $f_A=f_V$ model, and back-to-back with either the lepton or neutrino in the $f_A=0$ model.  Thus, we require $(\cos\theta_{\g\ell}-1)^2 +(\cos\theta_{\g\nu}+1)^2/3 >0.4$ or $(\cos\theta_{\g\nu}-1)^2 +(\cos\theta_{\g\ell}+1)^2/3 >0.4$ for the $f_A=0$ model, and only the former relationship for $f_A=f_V$.  This reduces \eff in both modes and models by 40\%.  We observe 0 (0) data events in the electron (muon) mode with $\Nbkg = 0.6 \pm 0.1~(1.0 \pm 0.4)$ for the $f_A=f_V$ model, corresponding to $\BR(\lnugam)<3.0\times10^{-6}$.  Likewise, in the $f_A=0$ model, we observe 3 (2) data events with $\Nbkg = 1.2 \pm 0.4~(1.5 \pm 0.6)$, corresponding to $\BR(\lnugam)<18\times10^{-6}$.

\bigskip
\section{Conclusion}
In conclusion, we have searched for \lnugam using a hadronic recoil technique and observe no significant signal within a data sample of 465 million \BB pairs.  We report a model-independent limit of $\BR(\lnugam)<15.6\times10^{-6}$ at the 90\% CL, which is consistent with the standard model prediction and is the most stringent published upper limit to date.  Using Eq.~\eqref{lnugamBF} with $f_B = 0.216\pm0.022\gev$, $m_B=5.279\gevcc$, $\tau_B=1.638\,\ps$, $m_b = 4.20\gevcc$, and $\lvert V_{ub}\rvert = (3.93\pm0.36)\times10^{-3}$ \cite{ref:pdg}, the combined branching fraction likelihood function corresponds to a limit of $\lambda_B> 0.3\gev$ at the 90\% CL.


\bigskip 

\begin{thebibliography}{99}   


\bibitem{ref:burdman} G.\ Burdman, T.\ Goldman, and D.\ Wyler, Phys.\ Rev.\ D {\bf51}, 111 (1995).
\bibitem{ref:superB} SuperB Collaboration, SuperB Conceptual Design Report, arXiv:0709.0451v2 [hep-ex], (2007).
\bibitem{ref:cleo} CLEO Collaboration, T.\ E.\ Browder {\em et al.}, Phys.\ Rev. D {\bf56}, 11 (1997).
\bibitem{ref:KPY} G.\ P.\ Korchemsky, D.\ Pirjol, and T.\ M.\ Yan, Phys.\ Rev.\ D {\bf61}, 114510 (2000).
\bibitem{ref:Lunghi} E.\ Lunghi, D.\ Pirjol, and D.\ Wyler, Nucl.\ Phys.\ B {\bf649}, 349 (2003);
S.\ Descotes-Genon and C.\ T.\ Sachrajda, Nucl.\ Phys.\ B {\bf650}, 356 (2003).
\bibitem{ref:yaouanc} A.\ Le Yaouanc, L.\ Oliver, and J.-C. Raynal, Phys.\ Rev.\ D {\bf77}, 034005 (2008).
\bibitem{ref:ball} P.\ Ball and E.\ Kou, JHEP {\bf0304}, 29 (2003).
\bibitem{ref:kou} D.\ Becirevic, B.\ Haas, and E.\ Kou, arXiv:0907.1845v1 [hep-ph], (2009).
\bibitem{ref:babar} \babar\ Collaboration, B.\ Aubert {\em et al.}, Nucl.\ Instr.\ Meth. A {\bf479}, 1 (2002).
\bibitem{ref:mine} \babar\ Collaboration, B.\ Aubert {\em et al.}, arXiv:0907.1681 [hep-ex], submitted to Phys.\ Rev.\ D (RC), (2009).
\bibitem{ref:punzi} G.\ Punzi, 	arXiv:physics/0308063v2 [physics.data-an], (2003).
\bibitem{ref:latMom} ARGUS Collaboration, A.\ Drescher {\em et al.}, Nucl.\ Instr.\ Meth. A {\bf237}, 464 (1985).
\bibitem{ref:FC} G.\ J.\ Feldman and R.\ D.\ Cousins, Phys.\ Rev.\ D {\bf57} 3873 (1998).
\bibitem{ref:pdg} Particle Data Group, C.\ Amsler {\em et al.}, Phys.\ Lett. B {\bf667}, 1 (2008).


\end{thebibliography}

\end{document}